\documentstyle[emulateapj,flushrt]{article}
\newcommand{\beq}{\begin{equation}}
\newcommand{\eeq}{\end{equation}}
\newcommand{\bea}{\begin{eqnarray}}
\newcommand{\eea}{\end{eqnarray}}
\newcommand{\bml}{\begin{mathletters}}
\newcommand{\eml}{\end{mathletters}}
\newcommand{\pd}[3]{\frac{\partial^{#3}{#1}}{\partial {#2}^{#3}}}
\newcommand{\td}[3]{\frac{d^{#3}{#1}}{d{#2}^{#3}}}
\newcommand{\rem}[1]{ }
\begin{document}
\title{Axisymmetric Self-Similar Equilibria of Self-Gravitating Isothermal 
Systems}
\author{Mikhail V. Medvedev\altaffilmark{1} and Ramesh Narayan }
\affil{Harvard-Smithsonian Center for Astrophysics,
60 Garden St., Cambridge, MA 02138}
\altaffiltext{1}{Also at the Institute for Nuclear Fusion, RRC ``Kurchatov
Institute'', Moscow 123182, Russia; E-mail: mmedvedev@cfa.harvard.edu;
URL: http://cfa-www.harvard.edu/\~{ }mmedvede/ }

\begin{abstract}
All axisymmetric self-similar equilibria of self-gravitating, 
rotating, isothermal systems are identified by solving the nonlinear Poisson 
equation analytically. There are two families of equilibria: 
(1) Cylindrically symmetric solutions in which the density varies with
cylindrical radius as $R^{-\alpha}$, with $0\le\alpha\le2$.
(2) Axially symmetric solutions in which the density varies as
$f(\theta)/r^2$, where $r$ is the spherical radius and $\theta$ is the
co-latitude. The singular isothermal sphere is a special case of the latter
class with $f(\theta)={\rm constant}$. The axially symmetric equilibrium 
configurations form a two-parameter family of solutions and include 
equilibria which are surprisingly asymmetric with respect to the equatorial 
plane. The asymmetric equilibria are, however, not force-free at the singular 
points $r=0, \infty$, and their relevance to real systems is unclear. 
For each hydrodynamic equilibrium, we determine the 
phase-space distribution of the collisionless analog.
\end{abstract}

\section{Introduction}

Many astrophysical objects consist of equilibrium configurations in which
self-gravity is resisted by pressure (velocity dispersion) and centrifugal
forces. A particularly simple example, which is well-suited for practical
analysis, is the case of an isothermal fluid with a linear pressure--density 
relation, $p=\rho c_s^2$, where $p$ is the pressure, $\rho$ is the density,
and $c_s$ is the sound speed which is taken to be a constant. Equilibrium 
configurations of isothermal systems have been studied as models of galaxies
(\cite{Toomre82}; \cite{BT}; see also \cite{Richstone80}; \cite{Monetetal81} 
for some special cases) and newly formed stars
(e.g., \cite{Hayashietal82}; \cite{Kiguchietal87}).

Although some analytical solutions have been published previously, no 
systematic study of isothermal equilibria has been presented so far. In this 
paper we classify and derive analytically all possible self-similar 
axisymmetric equilibria of a self-gravitating isothermal system. We find 
that there are only two distinct classes of equilibria:
\begin{enumerate}
\item
Cylindrically symmetric equilibria, in which all quantities, such as density,
potential, velocity, etc., depend on the cylindrical radius, $R$, only;
\item
Axially symmetric equilibria, in which the quantities are functions of both the
spherical radius, $r$, and the co-latitude, $\theta$.
\end{enumerate} 

The cylindrically symmetric solutions are rather simple: the density varies as
$\rho\propto R^{-2(n+1)}$, and the azimuthal velocity as 
$\Omega R\equiv v_\varphi=v_{\varphi0}R^{-n}$, where the allowed range of 
$n$ is $-1\le n\le 0$.

The axially symmetric solutions are more rich. First, the density  
always varies as $\rho\propto 1/r^2$ and the rotation curve is 
flat, $v_\varphi=v_{\varphi0}={\rm constant}$. Second, these equilibria form
a two-parameter family of solutions. One of the parameters, $A$, determines 
the rotation velocity, $A=2+v_{\varphi0}^2/c_s^2$, and the other parameter, 
$B$, controls the symmetry of the solutions with respect to the equatorial 
plane. For $B=1$, the solutions are up-down symmetric. These solutions have 
been previously obtained by Toomre (1982) and Hayashi et al. (1982) and include
the singular isothermal sphere ($A=2$) and the cold Mestel disk ($A\to\infty$) 
as limiting cases. 

The solutions with $B\not=1$ are asymmetric with respect to the equatorial
plane. This contradicts Lichtenstein's theorem (\cite{Lich}; \cite{Wavre}) 
which states that a barotropic, self-gravitating equilibrium must have a 
plane of symmetry perpendicular to the axis of rotation. The paradox is 
resolved by noting that the solutions with $B\not=1$ are not force-free at two 
singular points, $r=0$ and $r=\infty$, where Poisson's equation is 
ill-defined. External forces have to be applied to the matter at 
the singularities to hold the system in equilibrium. This is proved 
rigorously for the case without rotation, $A=2$, and is likely to be 
correct also for rotating solutions. 

The paper is organized as follows. In \S \ref{S:SSS}, we present a general
analysis of the problem and classify all possible self-similar equilibrium
configurations of self-gravitating isothermal systems. We derive cylindrically 
symmetric solutions of the equation of hydrostatic equilibrium in \S \ref{A0}. 
We identify a two-parameter family of axially symmetric solutions in 
\S \ref{S:ASS} and describe the properties of these solutions 
in \S \ref{S:PROP}. In \S \ref{S:A=2} we discuss the non-rotating case and in 
\S \ref{S:A=INFINITY} the thin disk limit. In \S \ref{S:PDF} we determine the 
steady-state distribution function of collisionless stellar systems, and we 
summarize the conclusions in \S \ref{S:CONCL}.

\section{General Considerations and Self-Similar Solutions \label{S:SSS}}

We perform all calculations in spherical coordinates $(r,\theta,\varphi)$, 
but we also occasionally consider the cylindrical radius, $R=r\sin\theta$.

A hydrostatic equilibrium configuration satisfies the momentum equation with
vanishing time derivative, and Poisson's equation. By definition, the 
$\hat r$- and $\hat\theta$-components (``hat'' denotes unit vectors) of the 
velocity vanish in equilibrium. Also, by the condition of axial symmetry, all 
derivatives with respect to the toroidal angle, $\varphi$, vanish. Thus, 
we need to consider only the $\hat{r}$- and $\hat{\theta}$-components 
of the momentum equation and Poisson's equation for the potential:
\bml
\bea
\frac{v_\varphi^2}{r}&=&\frac{c_s^2}{\rho}\pd{\rho}{r}{}+\pd{\phi}{r}{} ,
\label{mom-r}\\
\frac{v_\varphi^2\cot\theta}{r}&=&\frac{c_s^2}{\rho r}\pd{\rho}{\theta}{}+
\frac{1}{r}\pd{\phi}{\theta}{} ,
\label{mom-theta}\\
\nabla^2_{r,\theta}\;\phi&=&4\pi G\rho ,
\label{Poisson}
\eea
\eml
where 
$$
\nabla^2_{r,\theta}=\frac{1}{r^2}\pd{}{r}{}\,r^2\pd{}{r}{}+
\frac{1}{r^2\sin\theta}\pd{}{\theta}{}\,\sin\theta\pd{}{\theta}{}
$$
is the axisymmetric Laplacian operator,
$\phi$ is the gravitational potential, $v_\varphi$ is the toroidal 
component of the velocity due to rotation which is, in general, a function 
of $r$ and $\theta$, and $G$ is Newton's constant. The radial coordinate and
the density are taken to be dimensionless throughout the paper. 

\subsection{Equilibrium of an Isothermal Gas}

We begin with a general treatment of the problem, without any assumption of 
self-similarity. We take the gas to be isothermal, i.e.,
the pressure and the density are related as follows,
\beq
p=c_s^2\rho ,
\label{p}
\eeq
where $c_s={\rm const.}$ is the sound speed. Since an isothermal
system is a special case of a barotropic system, $p=p(\rho)$,
the following generic properties immediately follow from the Poincar\'e-Wavre
theorem (\cite{Tassoul}): (i) the angular velocity is constant on
cylinders centered on the axis of rotation, i.e., in cylindrical
coordinates $(R,\varphi,z)$, the velocity $v_\varphi=v_\varphi(R)$ is
independent of $z$; (ii) the effective gravity can be derived from an effective
potential; and (iii) the effective gravity is normal to the isopycnic (i.e.,
constant density) surfaces. Tassoul (1978) has derived general relations 
between $\rho,\ \phi$, and $v_\varphi$ in a rotating barotropic system;
we briefly re-derive some of these results for completeness.

Eliminating $v_\varphi$ between 
Eqs. (\ref{mom-r},b), we arrive at the following differential equation:
\beq
\left(\pd{}{\,\ln r}{}-\pd{}{\,\ln\sin\theta}{}\right)
\left(c_s^2\ln\rho+\phi\right)=0 .
\eeq
Any function of the argument $\ln(r\sin\theta)$ is a solution of this 
differential equation. Absorbing the logarithm into the definition of
the function, we write the solution as follows
\beq
\phi(r,\theta)=-c_s^2\ln\rho(r,\theta)+u(r\sin\theta) ,
\label{phi}
\eeq
where $u$ is an arbitrary function of the cylindrical radius $R=r\sin\theta$.
The function $u$ is related to the rotation velocity as follows,
\beq
v^2_\varphi(r,\theta)=v^2_\varphi(R)=R\,\pd{\,u(R)}{\,R}{} .
\label{v-phi}
\eeq
This relation is obtained upon substituting Eq.\ (\ref{phi}) 
into Eq.\ (\ref{mom-r}).

Next, we use Poisson's equation to relate the density and gravitational
potential. Substituting Eq.\ (\ref{phi}) into Eq.\ (\ref{Poisson})
and using  Eq.\ (\ref{v-phi}) and the identity
$$
\nabla^2_{r,\theta}\,u(r\sin\theta)
=\nabla_R^2\,u(R)
=\frac{1}{R}\pd{}{R}{}\left(R\pd{}{R}{}\,u(R)\right),
$$ 
we obtain the following equation for the density distribution,
\beq
-c_s^2\nabla^2_{r,\theta}\,\ln\rho+\frac{1}{R}\pd{}{R}{}\,v^2_\varphi
=4\pi G\rho .
\label{main1}
\eeq
Given the rotation curve, $v_\varphi(R)$, the solution of Eq.\ (\ref{main1}), 
together with Eqs.\ (\ref{p}),(\ref{phi}), and (\ref{v-phi}), completely 
determines the solution. To proceed further, we need to make some simplifying 
assumptions.

\subsection{Self-Similar Solutions}

We now look for self-similar solutions of Eq.\ (\ref{main1}). Let us assume
that $v_\varphi$ is described by a power-law in $R=r\sin\theta$. Then the 
density distribution is also a power-law in $r$, and we write
\beq
\rho=\rho_0\frac{f(\theta)}{r^\alpha},\qquad 
v_\varphi=\frac{v_{\varphi 0}}{R^n} ,
\label{try-sol}
\eeq
where $f(\theta)$ is an unknown function to be calculated, $\rho_0$ and
$v_{\varphi 0}$ are normalization constants, and $\alpha$ and $n$ are
parameters. Substituting Eqs.\ (\ref{try-sol}) into Eq.\ (\ref{main1}), 
we obtain
\beq
\frac{c_s^2}{r^2}\left(\alpha-\frac{1}{\sin\theta}\pd{}{\theta}{}\;
\sin\theta\pd{}{\theta}{}\,\ln f(\theta)\right)
-\frac{2n\, v_{\varphi 0}^2}{(r\sin\theta)^{2n+2}}
=4\pi G\rho_0\frac{f(\theta)}{r^\alpha} .
\label{main2}
\eeq
This equation can be satisfied only when the powers of $r$ on the various
terms match. There are two possible cases:
\begin{enumerate}
\item
Cylindrically symmetric solutions. In this case, the first term on the 
left-hand-side of Eq.\ (\ref{main2}) vanishes identically and $2n+2=\alpha$.
\item
Axially symmetric solutions. In this case, $\alpha=2$ and $n=0$ and 
the term proportional to $v_{\varphi 0}$ in Eq.\ (\ref{main2}) vanishes.
These solutions have flat rotation curves.
\end{enumerate}
The above list exhausts all possible cases. Thus, there are only two families
of self-similar, axisymmetric equilibrium configurations of self-gravitating 
isothermal systems with permanent rotation. 

We now describe the two families of solutions in detail. In \S \ref{A0} 
we discuss the cylindrical solutions, and in \S\S 
\ref{S:ASS}--\ref{S:A=INFINITY} we discuss the axially symmetric solutions.

\section{Cylindrical solutions \label{A0} }

In this case, the first term on the left-hand-side of Eq.\ (\ref{main2}) 
vanishes identically, so that $2n+2=\alpha$. This condition forces $f$ to be 
$f(\theta)=(\sin\theta)^{-\alpha}$. The self-similar solution for $\rho$ 
in terms of $R=r\sin\theta$ is easily obtained from Eq.\ (\ref{main2}),
\bml
\beq
\rho(R)=-\frac{n\,v^2_{\varphi 0}}{2\pi G}
\frac{1}{R^{2n+2}}.
\label{cyl-sol-rho}
\eeq
We notice that this solution is physical only for $n\le0$; otherwise,
the density is negative.
The gravitational potential is determined from Eq.\ (\ref{phi}). We have,
\beq
\phi=-c_s^2\ln\!\left(-\frac{n\,v^2_{\varphi 0}}{2\pi G}
\frac{1}{R^{2n+2}}\right)+ u(R)+\phi_0,
\label{cyl-sol-phi}
\eeq
\label{cyl-sol}
\eml
\noindent
where $\phi_0$ is a constant that defines the zero-level of the potential and 
the function $u(R)$ is obtained by integrating 
Eq.\ (\ref{v-phi}) for a given power-law rotation curve,
$$
u(R)=\left\{\begin{array}{ll}
v^2_{\varphi 0}/(2n+2)R^{2n+2}, & \textrm{if } n\not=0,\\
v^2_{\varphi 0}\ln(R), & \textrm{if } n=0.
\end{array}\right.
$$
The solution is well behaved everywhere except at the axial 
singularity,\footnote{
	The singularity of the solution at $\theta=0$ is due to the singularity 
	of the differential operator of Eq.\ (\ref{main2}) on the axis.} 
where a Dirac $\delta$-function, i.e., a central mass ``wire,''  may be 
located. This additional mass density
proportional to $\delta(R)$ may be calculated using Gauss' integral
theorem for the flux of the gravitational field through a surface,
\beq
\int_S (-\nabla\phi)\cdot{\bf n}\,{\rm d}s=-4\pi G M_s ,
\label{Gauss}
\eeq
where $M_s$ is the mass enclosed within a volume bounded by a surface $S$ and 
${\bf n}$ is the unit normal outward from the volume. We choose the surface $S$ 
to be an axial cylinder of radius $\Delta R$ and length $L$, centered at 
$R=0$. Since $\phi$ is a function of $R$, [cf., Eq.\ (\ref{cyl-sol-phi})], 
the gradient $\nabla\phi$ is radial and is equal to
\beq
\nabla\phi=\hat R\left[\frac{(2n+2)c_s^2}{\Delta R}
+\frac{v^2_{\varphi 0}}{\Delta R^{2n+1}}\right],
\eeq
where $\hat R$ is the unit vector along the cylindrical radius. Calculating the 
enclosed mass via Gauss' theorem and taking the limit $\Delta R\to 0$ 
(remember, $n\le0$), we obtain the following expression for the linear density 
of the ``central wire,''
\beq
\mu_s\equiv\pd{M_s}{L}{}=\left\{\begin{array}{ll}
(1-|n|)\,c_s^2/G, & \textrm{if } n<0,\\
(c_s^2+v_{\varphi 0}^2/2)/G, & \textrm{if } n=0.
\end{array}\right.
\eeq
Clearly, the mass at the singularity becomes negative for $|n|>1$.
Thus, the self-similar, cylindrically symmetric solutions exist only for
$n$ in the range $-1\le n\le0$. The magnitude of the rotation velocity
$v_{\varphi 0}$ is arbitrary. It is worthwhile to note that for $n=-1$, the
axial $\delta$-function singularity disappears $(\mu_s=0)$ and the solution,
Eqs.\ (\ref{cyl-sol}), is well behaved everywhere. This case corresponds to 
solid-body rotation, $\Omega(R)=v_\varphi/R={\rm constant}$.

\section{Axially Symmetric Solutions \label{S:ASS}}

The equation for $\rho$ [Eq.\ (\ref{main1}) with $\alpha=2$ and $n=0$] reads
\beq
2-\frac{1}{\sin\theta}\pd{}{\theta}{}
\left(\sin\theta\pd{}{\theta}{}\,\ln f(\theta)\right)
=\frac{4\pi G\rho_0}{c_s^2}f(\theta) .
\label{MAIN}
\eeq
As one can see, the velocity gradient term drops out. 
That is, the density distribution appears to be independent of the 
rate of rotation, in apparent contradiction to the Poincar\'e-Wavre theorem.
As we shall see from the careful analysis below, this is not the case.
 
\subsection{Regular Solution}

It is remarkable that the nonlinear, second-order differential 
equation (\ref{MAIN}) may be solved analytically and a general two-parameter 
family of self-similar solutions can be found in a closed and explicit form. 
All details of this computation are presented in Appendix \ref{A1}. 
The solution is 
\bml
\beq
\rho(r,\theta)=\frac{c_s^2}{2\pi G}\,\frac{A^2}{\left(r\sin\theta\right)^2}\,
\frac{B\tan^A(\theta/2)}{\left[1+B\tan^A(\theta/2)\right]^2} ,
\label{sph-sol-rho}
\eeq
where $A>0$ and $B>0$ are two free parameters. We have restricted 
$A$ to be positive, because a change of sign of $A$ is identical to the 
replacement: $B\to1/B$. The gravitational potential calculated from 
Eq.\ (\ref{phi}) may be written in the form
\beq
\phi(r,\theta)=-c_s^2\ln\!\left(\frac{\rho(r,\theta)}
{\left|r\sin\theta\right|^{v_{\varphi 0}^2/c_s^2}}\right)+\phi_0 .
\label{sph-sol-phi}
\eeq
\label{sph-sol}
\eml
The denominator in the logarithm is the contribution due to rotation,
$u(r\sin\theta)=v_{\varphi 0}^2\ln(r\sin\theta)+{\rm const.}$, as follows
from Eq.\ (\ref{v-phi}). 
The requirement of regularity of the solution, i.e., the continuity of 
$\rho$ and $\phi$ and their derivatives everywhere, constrains one of the
free parameters,
\beq
A=2+v_{\varphi 0}^2/c_s^2 \ge2
\label{A}
\eeq
(see \S \ref{SING} for more details).
The other free parameter, $B$, remains unconstrained. Note that the density
distribution depends on the rotation velocity through Eq.\ (\ref{A}), 
in agreement with the Poinca\'e-Wavre theorem. For the special case $B=1$, 
the solution (\ref{sph-sol-rho}) reduces to the solution obtained by 
Toomre (1982) and Hayashi et al. (1982).\footnote{
	The parameter $A\equiv2n+2$ in Toomre (1982) and
	$A\equiv2\gamma$ in Hayashi et al. (1982). }

\subsection{Axial Singularity and a Generalized Solution \label{SING} }

The solution (\ref{sph-sol}) is smooth and well-behaved in the domain 
$0<\theta<\pi$. However, the differential operator in Eq.\ (\ref{MAIN}) is 
ill-defined at $\theta=0$. Therefore, a singular solution, proportional
to the Dirac delta-function, $\delta(\theta)$, is allowed on the axis. 
This additional mass density may be calculated using Gauss' theorem, 
Eq.\ (\ref{Gauss}). As a surface of integration, we choose a segment of a cone 
of revolution centered on the axis consisting of the following pieces, 
\bea
S_1&=&\left\{r=r_0; \ 0\le\theta\le\Delta\theta\right\},  \nonumber\\
S_2&=&\left\{r_0\le r\le r_0+\Delta r;\ \theta=\Delta\theta\right\},\nonumber\\
S_3&=&\{r=r_0+\Delta r; \ 0\le\theta\le\Delta\theta\},  \nonumber
\eea
where $r_0,\ \Delta r$, and $\Delta\theta$ are constants. The unit normals 
to the three surface pieces are ${\bf n}_1=-\hat r,\ {\bf n}_2=\hat\theta$, 
and ${\bf n}_3=\hat r$, respectively. The gradient of the potential for 
small angles reads
\beq
\nabla\phi\simeq\left(\frac{\beta c_s^2}{r}; \
\frac{(\beta-A) c_s^2}{r\theta}; \ 0\right) ,
\eeq
where $\beta=2+v_{\varphi 0}^2/c_s^2$. Performing the integration over the
surface $S=S_1\oplus S_2\oplus S_3$, taking the limit $\Delta\theta\to0$, 
and noticing that the radial distance, $\Delta r$, at $\theta=0$ is the 
length along the axis, $\left.\Delta r\right|_{\theta=0}=L$, we obtain the 
linear mass density of a singular density distribution on the axis,
\beq
\mu_s\equiv\pd{M_s}{L}{}=\frac{c_s^2}{2G}\left(2
+\frac{v_{\varphi 0}^2}{c_s^2}-A\right) .
\label{mu-sing}
\eeq
The requirement of a non-singular density, $\mu_s=0$, gives back the relation
(\ref{A}). In the more general case, we see that $A$ is related to the
rotation velocity and the mass density at the axis. 
The mass on the axis cannot be negative, hence we have the constraint
\beq
A\le 2+ v_{\varphi 0}^2/c_s^2 .
\eeq

We note that the radial part of the spherical Laplacian, $\nabla^2_{r,\theta}$, 
is ill-defined at $r=0$, so that the solution may also contain a point mass,
$\delta(r)$, at the origin. However, using Gauss' theorem for a sphere of 
radius $\Delta r\to0$ enclosing the origin, we find no
singular mass can be  ``hidden'' at $r=0$.

In the rest of the paper we consider only the regular solutions with 
$\mu_{\rm s}=0$. These solutions satisfy $A=2+v_{\varphi 0}^2/c_s^2$
[equation (\ref{A})].

\section{Properties of the Axially Symmetric Solutions \label{S:PROP} }

\subsection{General Properties}

Figures\ \ref{f:shape-a}---\ref{f:shape-d} show typical examples of the 
self-similar solution (\ref{sph-sol}). Contours of equal density for 
different values of the parameters $A$ and $B$ are displayed. 
As is seen, the parameter $A$ determines the overall shape of the matter 
distribution. As $A\to0$ the configuration tends to the cylindrically 
symmetric limit, and an axial singularity with $\mu_s\not=0$ is always present
[see Eq.\ (\ref{mu-sing})]. Figure \ref{f:shape-a} shows an example with
$A=0.7$. The density does not vanish at infinity at $\theta=0,\ \pi$.
Solutions without an axial singularity, $\mu_s=0$, exist for $A\ge2$, 
cf., Eq.\ (\ref{A}). The non-rotating limit, $A=2$, yields confocal 
ellipsoids/spheres, as shown in Fig.\ \ref{f:shape-b} (see \S \ref{S:A=2}).
For $A>2$, toroidal configurations with rotation are obtained, as shown in 
Figs.\ \ref{f:shape-c} and \ref{f:shape-d}. Interestingly, as $A\to\infty$, the
profile flattens and tends to a thin disk configuration, as shown in Fig.\
\ref{f:shape-d} for $A=20$. 

The parameter $B$ is responsible for up-down asymmetry. For $B=1$, the 
solutions are symmetric with respect to the equatorial plane. The solutions 
with $B>1$ are shifted (distorted) upwards, those with $B<1$ are shifted 
downwards. Note that solutions with $B$ and $1/B$ are identical except
that they are turned upside-down with respect to each other. Note also that 
the solutions with $B\not=1$ violate Lichtenstein's theorem (\cite{Lich}; 
\cite{Wavre}), which proves the existence of an equatorial plane of symmetry.
The resolution of this paradox is discussed below (\S \ref{S:A=2}).

\subsection{The Finite System Limit}

We now determine some global properties, such as the total mass, $M$, 
gravitational  energy, $W$, and angular momentum, ${\bf L}$, of the axially 
symmetric self-similar solutions as functions of $A$ and $B$. These 
quantities diverge unless a cutoff is imposed at large radii. The most 
natural way to do this is to assume that the system is immersed in a medium 
with finite pressure, $p_{\rm ext}$, but vanishing density. As follows 
from Eq.\ (\ref{p}), the cutoff surface in this case is the iso-density 
surface with $\rho_c=p_{\rm ext}/c_s^2$. The equation for the cutoff surface is
\beq
r_c(\theta)=\sqrt{g\,\Theta(\theta)/\rho_c},
\label{r-c}
\eeq
where $g=c_s^2/(2\pi G)$ and
$$
\Theta(\theta)=\frac{A^2}{\sin^2\theta}\,
\frac{B\tan^A(\theta/2)}{\left[1+B\tan^A(\theta/2)\right]^2} 
$$
is the angular part of the function, $\rho(r,\theta)$.
Notice that if $\mu_s\not=0$,\ $M$ and $W$ still diverge due to the
infinite mass located on the axis. So, we set $\mu_s=0$, which means that we
are restricted to the solutions with $A\ge2$.
The total mass, gravitational potential energy, and angular momentum
are defined as follows,
\beq
M=\int_V\rho\,{\rm d}V, \quad
W=\frac{1}{2}\int_V\rho\,\phi\;{\rm d}V, \quad
{\bf L}=\int_V\rho\,{\bf v\times r}\;{\rm d}V.
\label{defs}
\eeq
The binding energy is then $E=-(W+K)$, where the kinetic energy is
$K=Mv_{\varphi0}^2/2$. The surfaces of constant density and constant potential 
do not coincide due to rotation, unless $A=2$. For this reason, the potential
at the cutoff surface is a function of position, 
$\phi_c=\phi\left(r_c(\theta),\theta\right)$. To get rid of insignificant
constants in the expression for $W$, we redefine the gravitational potential
such that the maximum potential at the surface is zero. This constrains
the constant $\phi_0$. The value of $\theta_{\rm m}$, the extremum point 
of $\phi$ along the $r_c(\theta)$, is found to satisfy the equation, 
$\tan^A(\theta_{\rm m}/2)=1/B$. Then, we have
\beq
\phi_0=
c_s^2\ln\rho_c-v_{\varphi 0}^2\ln\sqrt{g/\rho_c}
-v_{\varphi 0}^2\ln(A/2).
\eeq
As shown in \S \ref{S:A=2}, the solutions with $B\not=1$ are not
force-free; they include an external gravitational potential gradient. 
We do not include the
external potential in our definition of the gravitational energy, $W$.
We now calculate all the quantities (note, $L_z$ is the only 
nonvanishing component of the angular momentum). The details of the 
calculation are given in Appendix \ref{A2}. The results are
\bml
\bea
M&=&\frac{2\pi g^{3/2}}{\sqrt{\rho_c}}\,
\frac{\pi(A^2-4)}{16\cos(\pi/A)}\left(B^{1/A}+B^{-1/A}\right),\\
W&=&-\frac{\pi g^{3/2}c_s^2}{\sqrt{\rho_c}}\,
\frac{\pi(A-2)^2}{16\cos(\pi/A)}\Biggl(2+(A+2)\Biggl[\frac{1}{2}-\ln 2-\gamma 
\nonumber \\ & &{ }  
+\frac{\pi}{2}\,\tan\frac{\pi}{A}-
\psi\!\left(\frac{1}{2}+\frac{1}{A}\right)
\Biggr]\Biggr)
\left(B^{1/A}+B^{-1/A}\right), \\
L_z&=&\frac{\pi g^2c_s}{\rho_c}\,
\frac{\pi(A^2-1)}{12\sin(\pi/A)}\sqrt{A-2}\left(B^{1/A}+B^{-1/A}\right),
~~~~~~~~~~~
\eea
\label{mwl}\noindent 
\eml
\noindent
where $g=c_s^2/(2\pi G)$,\  $\psi(x)=\Gamma'(x)/\Gamma(x)$ is the digamma 
function, and $\gamma\equiv-\psi(1)=0.57721\ldots$ is Euler's constant.
The dependences of $M, \ W$ and $L_z$ as functions of $A$ are plotted in 
Fig.\ \ref{f:MWL}. All these increase with increasing $A$, i.e., as the 
rotation velocity $v_{\varphi0}$ increases. The dependence of these quantities 
as a function of $B$ is trivial; all are proportional to 
$\left(B^{1/A}+B^{-1/A}\right)$ and have minima at $B=1$, 
the symmetric configuration.

We emphasize that the truncation of a system at a finite $\rho=\rho_c$ breaks 
the self-similarity of the solutions. Thus, the above equations are approximate
and should be used with care. Nevertheless, Eqs.\ (\ref{mwl}) suggest that
both the specific angular momentum of a finite configuration, $L_z/M$, and the
gravitational binding energy per unit mass, $W/M$, are functions only
of $A$ and independent of $B$. The kinetic energy per unit mass,
$v_{\varphi0}^2/2$, is also a function only of $A$. Finally, if we consider the
specific entropy, $s=S/M$, where $S\propto-\int_V\rho\ln\rho\,{\rm d}V$, we see
that $S$ differs from the gravitational energy by additive and multiplicative
constants only. Thus $s$ is again independent of $B$. The question then is: if
energy and entropy are indistinguishable and independent of $B$, what
determines which value of $B$ is selected by a finite isothermal system with a
given value of $L_z/M$? The answer, as we show in the next section, is the
boundary condition, namely, the magnitude of external forces acting on the mass.

\section{The Nonrotating Limit, $A=2$ \label{S:A=2} }

We first prove that nonrotating solutions, $A=2$, consist of nested confocal 
ellipsoids. The solution (\ref{sph-sol-rho}) reads
\beq
\rho(r,\theta)=\frac{2c_s^2}{\pi G}\,\frac{1}{\left(r\sin\theta\right)^2}\,
\frac{B\tan^2(\theta/2)}{\left[1+B\tan^2(\theta/2)\right]^2} .
\label{ellips}
\eeq
The equation of an iso-density contour line is obtained 
by setting $\rho(r,\theta)=\rho_1={\rm constant}$. Upon
straightforward trigonometric simplifications, we obtain
\bea
r(\theta)&=&\left(\frac{2c_s^2B}{\pi G\rho_1(1+B)^2}\right)^{1/2}
\left[1+\left(\frac{1-B}{1+B}\right)\cos\theta\right]^{-1} \nonumber\\
&=&\frac{a(1-\epsilon^2)}{1+\epsilon\cos\theta},
\eea
This is the equation of an ellipse with eccentricity, $\epsilon$, and  
semi-major axis, $a$, with the origin located at one of the focii. 
The three-dimensional iso-density surfaces are 
the surfaces of revolution about the major axis.
Note that the case $B=1$ corresponds to the singular isothermal sphere.

Next, we demonstrate that the solutions with $B\not=1$ are not force-free. 
Let us consider an equidensity (which is also an equipotential) surface with 
density $\rho_1$. The gravitational potential inside the surface 
produced by the ``outside'' mass, $\rho<\rho_1$, is linear in the 
vertical coordinate, $z$, as shown in Appendix \ref{A3}:
\bea
\Phi&=&-z\;\sqrt{2\pi G c_s^2}\,\sqrt{\rho_1}\left(B^{1/2}-B^{-1/2}\right)
\nonumber \\
& &{ }\quad\times
\frac{1-\epsilon^2}{\epsilon^2}\left[\frac{1}{2|\epsilon|}
\ln\!\left(\frac{1+|\epsilon|}{1-|\epsilon|}\right)-1\right].
\label{extra-pot}
\eea
As one can see, $\Phi\propto\sqrt{\rho_1}$. On the other hand, the mass inside 
the equipotential is 
$M=\pi g^{3/2}/\sqrt{\rho_1}\left(B^{1/2}+B^{-1/2}\right)$, i.e.,
$M\propto1/\sqrt{\rho_1}$, as follows from Eq.\ (\ref{mwl}). The total
gravitational force exerted by the ``outside'' mass on the ``inside'' mass,
${\bf F}_g=-M\,\nabla\Phi$, is, thus, independent of $\rho_1$, i.e.,
independent of the the equidensity surface chosen:
\beq
{\bf F}_g=-\hat z\;\frac{2c_s^4}{G}\;\frac{1}{\epsilon}
\left[\frac{1}{2|\epsilon|}
\ln\!\left(\frac{1+|\epsilon|}{1-|\epsilon|}\right)-1\right],
\label{force}
\eeq
where $\epsilon=-(B-1)/(B+1)$. Taking an equipotential surface infinitely 
close to the origin, one can see that there is a net finite gravitational 
pull, ${\bf F}_g$, (upwards for $B>1$ and downwards for $B<1$) acting on the 
matter at the singularity, $r=0$. This force is produced by the gravity of all
the mass of the system. For the system to remain in equilibrium, there must be 
an equal and opposite external force, $-{\bf F}_g$, applied at $r=0$. 
A similar consideration shows that there must also be 
an extra force acting at infinity or at the last equidensity surface of a
finite system. These external forces are uniquely related to the asymmetry 
parameter, $B$, and,
hence, determine the structure of the equilibrium. When $B=1$, the external
forces vanish and such an equilibrium configuration is force-free.
While this analysis is restricted to the non-rotating case, $A=2$, 
similar results should hold for any $A$. We have verified this numerically.

As we mentioned earlier, the asymmetric solutions apparently contradict 
Lichtenstein's theorem (\cite{Lich}; \cite{Wavre}), according to which
any rotating body for which the angular velocity does not depend on $z$ 
(which is true for a self-gravitating isothermal system) always possesses
an equatorial plane of symmetry which is perpendicular to the axis of rotation.
There are two assumptions (among others) used in the theorem which are violated 
by our $B\not=1$ solutions: (i) no exernal potential is allowed and 
(ii) the system is assumed
to be bounded and the density is nowhere infinite. As we have demonstrated, 
an asymmetric configuration with $B\not=1$ exists only when 
external forces act on the system and pull it apart, in violation of (i). 
These forces are applied at the positions where condition (ii) is violated:
either $\rho$ or $r$ is infinite, and the solution is ill-defined.

\section{The Thin Disk Limit, $A\to\infty$ \label{S:A=INFINITY} }

The solutions with $A>2$ have rotation. The 
systems become flattened as the rotation increases and they tend to a thin disk
as $A\simeq v_{\varphi0}^2/c_s^2\to\infty$. In these solutions
the parameter $B$ determines the 
``inclination angle.'' We define the inclination angle (or the opening
angle of the ``equatorial cone''), $\theta_m$, as the angle at which the
density is maximum for a fixed radial distance, $r$. In the large-$A$ limit, we 
can neglect the\ $\sin^2\theta$ in the denominator in Eq.\ (\ref{sph-sol-rho}).
We then find
\beq
\theta_m\simeq 2\;\textrm{arc\,cot}\left(B^{1/A}\right) .
\eeq
Taking into account that $\textrm{arc\,cot}(x)=\pi/2-\textrm{arc\,cot}(1/x)$ 
and expanding the solution
(\ref{sph-sol-rho}) about $\theta_m$, for small angles 
$\theta-\theta_m\equiv\vartheta\ll\textrm{arc\,cot}\!\left(B^{\pm1/A}\right)$, 
we obtain
\beq
\rho(r,\vartheta)\simeq 
\frac{c_s^2}{8\pi G}\,\frac{A^2}{r^2}\left(\frac{1+B^2}{2B}\right)^2
\textrm{sech}^2\!\left[\vartheta\,A\left(\frac{1+B^2}{2B}\right)\right] ,
\eeq
For large $A$, the\ $\textrm{sech}^2$ function is very sharply peaked. 
Thus, the height-to-radius ratio of the disk may be estimated as 
\beq
\frac{H}{R}\simeq\Delta\vartheta\simeq\frac{c_s^2}{v_{\varphi0}^2}
\left(\frac{2}{B+1/B}\right) .
\eeq
In the symmetric case, $B=1$, this reduces to the standard result, 
$H/R\sim c_s^2/v_{\varphi0}^2\ll1$. As $c_s\to 0$ the disk becomes the
cold Mestel disk having a flat rotation profile.

Toomre (1999, private communication) has studied the dynamics of a truncated 
cold thin ``conical'' disk with $A\to\infty,\ B\not=1$. Figure \ref{fig:toomre}
shows his results for the evolution of a conical disk which is at $t=0$ has a 
self-similar form calculated in this paper. No external forces are applied at 
either the inner edge or outer edge of the disk. Because of the unbalanced 
gravitational force acting on the innermost ring of the disk, the core region 
is accelerated upward. The system evolves as a function of time in a 
self-similar fashion, as shown by a sequence of curves in Figure 
\ref{fig:toomre} which corresponds to equally spaced times in $\log t$.
The initial conical disk is thus destroyed within a dynamical time, showing
that a tree conical disk is highly unstable. On the other hand, if a
downward external force of appropriate magnitude is applied on the innermost 
ring of the conical disk and an equal upward force is applied on the 
outermost ring, the system is fully stable.

\section{Equilibrium of Collisionless Stellar Systems \label{S:PDF} }

There is a close analogy between fluid (gaseous) self-gravitating equilibrium
objects and stellar collisionless systems (cf., \cite{BT}). 
The analogy arises because a fluid system is supported against gravity by 
pressure gradients, while a stellar system is supported by gradients in the 
stress tensor which in many respects are like the pressure except that they 
can be anisotropic. There is a unique connection between the density and
velocity fields of 
fluid equilibrium configurations and the distribution functions of
stars in their collisionless analogs. For any stellar distribution function,
$f$, the profiles of density, streaming velocity, $\bar v_\varphi$, velocity 
dispersion, etc., are calculated uniquely by taking the moments of $f$. 
Similarly, given $\rho,\ \bar v_\varphi$, etc., an equilibrium $f({\bf x,v})$ 
may be determined. In general, there is no guarantee that the stellar 
distribution function so obtained will be physical, i.e., nonnegative over the 
entire six-dimensional phase space (see, e.g., \cite{BT} for more 
discussion). Remarkably, however, the distribution finction is guaranteed
to be positive for isothermal systems: the structure of an equilibrium of a
self-gravitating isothermal gas is identical to the structure of a
collisionless system of stars. We now determine the distribution function 
of a stellar system which corresponds to the hydrostatic equilibria found 
in \S \ref{S:ASS}.

We consider here the simple case when the distribution function 
depends on two classical integrals of motion,
the energy and axial component of the angular momentum.\footnote{
	According to Jeans' theorem, the distribution function of a steady
	state, self-gravitating system may be presumed to be a function of at
	most three isolating integrals. Usually, two of the integrals are the 
	energy and axial angular momentum, while, in general, there is no 
	simple analytical form for the third integral (see, \cite{BT} for 
	discussion). }
Lynden-Bell (1962), and subsequently Hunter (1975), developed a method to derive
distribution functions from the density and mean velocity profiles.
We introduce the integrals of energy and angular momentum of a particle as
follows,
\beq
{\cal E}=\left(v_r^2+v_\theta^2+v_\varphi^2\right)/2+\phi(r,\theta)-\phi_0, 
\qquad {\cal L}_z=\left|v_\varphi\right|\,r\sin\theta ,
\eeq
where $v_i$ are the components of the particle velocity. Since the density
profile is insensitive to the direction of rotation, it yields, upon inversion,
a distribution function which is even in ${\cal L}_z$, i.e., 
$f_+=f({\cal E,L}_z)+f({\cal E,-L}_z)$. Thus, there are, in principle, 
infinitely many distributions which produce identical density profiles
and differ by an arbitrary, odd in ${\cal L}_z$, function, 
$f_-=f({\cal E,L}_z)-f({\cal E,-L}_z)$, determined by the sense of motion of
individual stars. For real stellar systems, $f_-$ can also be determined, 
given the average rotation profile, $\bar v_\varphi$. Henceforth, we focus on
the symmetric part of the distribution function. We omit the subscript
``$+$,'' since this should not cause any confusion.

To determine $f$, we need to express the density profile as a function 
of the cylindrical radius and gravitational potential. 
From Eqs.\ (\ref{sph-sol-phi}) and (\ref{A}), it follows that
\beq
\rho(R,\phi)=R^{A-2}\exp\left[-\left(\phi-\phi_0\right)/c_s^2\right] .
\eeq
We now observe that $\rho(R,\phi)$ is independent of $B$. Hence 
$f({\cal E,L}_z)$ derived from it is also independent of $B$.\footnote{
	More precisely, it depends on $B$ via $\phi(r,\theta)$ entering 
	the definition of ${\cal E}$.}
Since $f({\cal E,L}_z)$ is unique (up to $f_-$), the distribution function
proposed by Toomre (1982) as the starting point of his analysis for
the specific case of $B=1$, is, in fact, valid for any value of $B$
and reads,
\beq
f({\cal E,L}_z)=f_0\,{\cal L}_z^{A-2}\exp\left(-{\cal E}/c_s^2\right),
\eeq
where $f_0$ is a normalization constant.
We can now write the phase-space distribution function of a steady-state,
self-gravitating system of stars, $f({\bf x,v})$, in an explicit form
as follows,
\beq
f(r,\theta,{\bf v})=f_0\,\rho(r,\theta)\,
\exp\!\left(-\frac{v_r^2+v_\theta^2+v_\varphi^2}{2\,c_s^2}+
\frac{v_{\varphi0}^2}{c_s^2}\,\ln\left|v_\varphi\right|\right).
\eeq
Here $c_s^2\equiv\overline{v^2}-\overline{v}^2$ is the velocity dispersion of
stars and $|v_{\varphi0}|$ is their mean circular velocity. We again notice a
remarkable fact: the parameter $B$ defines the shape of the
gravitational potential, but does not affect the velocity distribution of 
stars in this potential.

\section{Conclusion \label{S:CONCL} }

In this paper, we analytically calculated all possible self-similar, 
axisymmetric equilibrium states of a self-gravitating, isothermal gas with 
rotation. We showed that there are two distinct classes of hydrostatic 
equilibria; namely cylindrically symmetric equilibria and axially symmetric 
equilibria. The axially symmetric solutions are more physical, since the 
matter density vanishes at infinity. Among 
the axially symmetric solutions, we found equilibrium states which are 
asymmetric with respect to the equatorial plane. Such states satisfy Poisson's 
equation, but are not force-free at the singularities, $r=0,\infty$.
It is the external forces that support the asymmetric shape.
This example shows that self-similar solutions should be treated with 
caution and checked for possibly unphysical boundary conditions that may be 
``hidden'' at their singular points. 

Are the asymmetric configurations likely to be realized in nature? 
Since real systems are finite, we should truncate our solutions at 
some $\rho_c$ and throw away the ``outside'' mass. For a truncated solution 
to be valid, the gravitational potential of the discarded ``outside'' mass has 
to be replaced by a suitable external potential. As Eq.\ (\ref{extra-pot}) 
shows, the potential must have a linear gradient in $z$ (for a nonrotating
solution) of a magnitude determined by $\rho_c$ and $B$. Such a potential 
may be produced, for instance, by an external object located on the $z$ 
axis at a distance large compared to the size of the system. Second, 
to keep the system at rest, another force of equal magnitude and opposite 
sign must act on the central core. Since the force must act on the core alone,
it must be of non-gravitational origin, which, clearly, is not easy to
arrange. One could imagine, for example, that the core is ionized (by radiation 
of a central source, for instance), while the material outside the core is 
neutral. Then, if an external magnetic field threads the ionized core, one 
could imagine the field pinning the core, but having no influence on the 
neutral outer material. The distortion from equatorial symmetry would then 
be determined by the strength of the gravitational attraction of the external 
object and the counterbalancing force from the tension and curvature of the
field lines. Although this construction is rather artificial, it demonstrates 
that asymmetric, non-force-free equilibria may, at least in principle, exist 
in nature. If they do, and if the systems are isothermal, they will resemble
some of the $B\not=1$ solutions derived in this paper.

\acknowledgments

We are very grateful to Alar Toomre for several  valuable contributions to this 
paper. In particular, he suggested that all asymmetric equilibria must be 
unstable in the absence of external forces and provided the computer code
and numerical results presented in Figure \ref{fig:toomre}. He also suggested
the technique used in Appendix \ref{A3} to calculate the potential inside an 
asymmetric ellipsoidal equilibrium. We also thank George Rybicki for useful 
discussions and Kristen Menou for help with the translation from French of 
the proof of Lichtenstein's theorem given by Wavre (1932). 
This work was supported in part by NASA grant NAG~5-2837
and NSF grant AST~9820686.

\begin{appendix}
\section{Method of solution of Eq.\ (\ref{MAIN}) \label{A1} }

Here we demonstrate how a general solution of Eq.\ (\ref{MAIN}) can be found
analytically. The equation we analyze reads
\beq
2-\frac{1}{\sin\theta}\td{}{\theta}{}
\left(\sin\theta\td{}{\theta}{}\,\ln f(\theta)\right)
=\kappa f(\theta) ,
\label{a1}
\eeq
where $\kappa={4\pi G\rho_0}/{c_s^2}$. First of all, we get rid of the
constant term on the left-hand-side. We observe that
\beq
\frac{1}{\sin\theta}\td{}{\theta}{}
\left(\sin\theta\td{}{\theta}{}\,\ln(\sin^\beta\theta)\right)=-\beta,
\eeq 
where $\beta$ is a constant. Thus, upon introducing a new function
$g(\theta)=f(\theta)/\sin^2\theta$, equation (\ref{a1}) becomes
\beq
\sin\theta\td{}{\theta}{}\left(\sin\theta\td{}{\theta}{}\,\ln g(\theta)\right)
+\kappa g(\theta)=0 .
\eeq
We now observe that\ $\sin\theta$ can be put into the denominator as follows,
$\partial\theta/\sin\theta=\partial\ln\left|\tan(\theta/2)\right|$. Let us 
redefine again the function $g(\theta)$ and change the independent variable 
$\theta$ as follows,
\beq
\xi=\ln\left|\tan\frac{\theta}{2}\right|, \qquad
w(\xi)=\ln{g(\theta)}=\ln\!\left(\frac{f(\theta)}{\sin^2\theta}\right) ,
\label{subs1}
\eeq
Then the differential equation is greatly simplified and reads\footnote{
	Note that the above substitutions are equivalent to Howard's
	transformation used by Toomre (1982). }
\beq
\td{}{\xi}{2}\,w(\xi)+\kappa\,e^{w(\xi)}=0 .
\label{Spitzer}
\eeq

One can now reduce this equation to a first-order differential equation
which can be integrated. Since Eq.\ (\ref{Spitzer}) contains no explicit
dependence on $\xi$, we introduce a {\em new} function $p$ 
which is a function of the {\em old} function $w$ as follows,
\beq
p(w)=\td{w(\xi)}{\xi}{}, \textrm{  so that  } 
\td{w(\xi)}{\xi}{2}=\td{p(w)}{\xi}{}=\td{p(w)}{w}{}\td{w(\xi)}{\xi}{}
=p(w)\td{p(w)}{w}{}.
\eeq 
Upon this substitution, the resulting differential equation, 
$p\,p'_w=-\kappa\,e^w$, is readily integrated for $p(w)$ to yield
\beq
\td{w(\xi)}{\xi}{}\equiv p(w)=\pm\sqrt{C_1-2\kappa\,e^{w(\xi)}} .
\label{tmp}
\eeq
Here $C_1$ is a first constant of integration which we assume, to be specific 
at the moment, to be positive (see discussion below). Making use of another
substitution,
\beq
v(\xi)=e^{w(\xi)}, 
\label{subs2}
\eeq
the resulting differential equation, $v'_\xi=\pm v\sqrt{C_1-2\kappa v}$,
can be integrated again. We arrive at the
expression:
\beq
\frac{1}{\sqrt{C_1}}\,\ln\!\left|\frac{\sqrt{C_1-2\kappa v(\xi)}-\sqrt{C_1}}
{\sqrt{C_1-2\kappa v(\xi)}+\sqrt{C_1}}\right| = \pm(\xi-C_2) ,
\eeq
where $C_2$ is a second constant of integration. We now resolve this
equation with respect to $v(\xi)$. There are two solutions, namely
\beq
v(\xi)=\pm\frac{2C_1}{\kappa}\frac{e_*(1\mp e_*)^2}{(1-e_*^2)^2} ,
\label{2sol}
\eeq
where we used the short-hand notation 
$e_*=\exp\left(\pm\sqrt{C_1}\left[\xi-C_2\right]\right)$. Comparing 
Eqs.\ (\ref{subs1}) and (\ref{subs2}), one can see that the density 
$\rho(r,\theta)\propto f(\theta)\propto v(\xi)$. Thus, for the density to be 
positive, the function $v(\xi)$ must be {\em real} and {\em positive}, also. 
This condition is satisfied for the upper sign in Eq.\ (\ref{2sol}). 
Thus, we have
\beq
v(\xi)=\frac{2C_1}{\kappa}\frac{e_*}{(1+e_*)^2}
=\frac{C_1/2\kappa}{\cosh^2\!\left(\sqrt{C_1}\left[\xi-C_2\right]/2\right)} .
\label{v-sol}
\eeq
Recalling all the definitions we made throughout the calculation, see Eqs.\ 
(\ref{subs1}) and (\ref{subs2}), we finally obtain the general solution
of the differential equation (\ref{MAIN}),
\beq
f(\theta)=\frac{C_1/\left(2\kappa\sin^2\theta\right)}{\cosh^2\!\left(\sqrt{C_1}
\left[\ln\left|\tan(\theta/2)\right|-C_2\right]/2\right)} .
\label{f-sol}
\eeq
Clearly, since $\cosh x$ has no zeros for $-\infty<x<\infty$, 
this solution is well behaved for all $\theta$'s from the range:
$0<\theta<\pi$. Therefore, we investigate some properties of this solution 
in more details in the main text. Quite interestingly, solution (\ref{f-sol})
can be simplified further and finally reads as follows,
\beq
f(\theta)=\frac{2A^2}{\kappa\sin^2\theta}\,
\frac{B\left|\tan(\theta/2)\right|^A}{\left(1+B\left|\tan(\theta/2)\right|^A
\right)^2} ,
\label{soln}
\eeq
where $A=\sqrt{C_1}$ and $B=\exp\left(-\sqrt{C_1}C_2\right)>0$ are new
constants of integration. The ``absolute value'' signs may be omitted if
$0\le\theta\le\pi$, i.e., the tangent is non-negative.

It can be rigorously shown that the particular choice of $C_1>0$ and $C_2$ 
being a real number is unique, provided the solution to be physically 
meaningful. The derivation following Eq.\ (\ref{tmp}) holds for the 
complex-valued constants $C_1$ and $C_2$. Separating real and imaginary parts 
in Eq.\ (\ref{2sol}) and requiring $v$ to be a {\em real-valued} function for 
{\em all} real $\xi$, one can arrive, upon straightforward but cumbersome 
mathematical manipulations, at the conclusion that $C_1$ must be positive 
and $C_2$ may be complex, such that ${\rm Im}\,\sqrt{C_1}C_2=\pi k,\ 
k=0,1,2,\ldots$ and its real part is arbitrary. From this, it follows that 
$e_*>0$ for even $k$'s and $e_*<0$ for odd $k$'s. The additional constraint 
that $v$ must be positive uniquely selects the solution given by 
Eq.\ (\ref{v-sol}).

\section{Expressions for the mass, gravitational potential energy, 
and angular momentum \label{A2} }

From Eqs.\ (\ref{sph-sol}), using Eqs.\ (\ref{A}) and (\ref{r-c}), 
we straightforwardly calculate
\bml
\beq
M=\frac{2\pi g^{3/2}}{\sqrt{\rho_c}}\,{\cal I}, 
\eeq
\beq
W=-c_s^2\,\frac{\pi g^{3/2}}{\sqrt{\rho_c}}\,\left\{\left[
\ln\rho_c +2-\frac{\phi_0}{c_s^2}
-(A-2)\left(\ln\sqrt{g/\rho_c}-1\right)\right]{\cal I}
-\left(\frac{A}{2}-1\right){\cal J}\right\},
\eeq
\beq
L=\frac{\pi g^2}{\rho_c}\,v_{\varphi0}\,{\cal K} , 
\eeq
\label{MWL}
\eml
where $g=c_s^2/(2\pi G)$ and $\cal I$, $\cal J$, and ${\cal K}$ are the 
integrals:
\beq
{\cal I}=\int_0^\pi\left[\Theta(\theta)\right]^{3/2}\sin\theta\,{\rm d}\theta,
\qquad
{\cal J}=\int_0^\pi\left[\Theta(\theta)\right]^{3/2}\,
\ln\left[\Theta(\theta)\sin^2\theta\right]
\sin\theta\,{\rm d}\theta,
\qquad
{\cal K}=\int_0^\pi[\Theta(\theta)]^2\,\sin^2\theta\,{\rm d}\theta .
\eeq
Changing the integration variable to $z=B\tan^A(\theta/2)$, 
the above integrals reduce to
\bml
\bea
{\cal I}&=&\frac{A^2}{2}\int_0^\infty
\left(B^{1/A}z^{1/2-1/A}+B^{-1/A}z^{1/2+1/A}\right)
\frac{1}{(1+z)^3}\;{\rm d}z, \\
{\cal J}&=&\frac{A^2}{2}\int_0^\infty
\left(B^{1/A}z^{1/2-1/A}+B^{-1/A}z^{1/2+1/A}\right)\frac{1}{(1+z)^3}\; 
\ln\left(\frac{A^2 z}{(1+z)^2}\right)\;{\rm d}z, \\
{\cal K}&=&\frac{A^3}{2}\int_0^\infty
\left(B^{1/A}z^{1-1/A}+B^{-1/A}z^{1+1/A}\right)
\frac{1}{(1+z)^4}\;{\rm d}z.
\eea
\eml

The first integral, ${\cal I}$, is symbolically identical to 
$I=\int_0^\infty z^{\alpha-1} R(z)\, {\rm d}z$, where $R(z)$ is a rational
function. It can be evaluated in the complex plane. The branch cut is taken 
to be from 0 to $+\infty$ along the real axis. We choose the contour of 
integration which consists of four pieces: (i) a circular part around the 
branching point $z=0$ of radius $R_<\to 0$, (ii) a linear part, $L_+$, along 
the cut from above extending from $R_<$ to $R_>$, (iii) a circular part of 
radius $R_>\to\infty$, and (iv) a linear part, $L_-$, going along the cut from 
below and closing the contour. One can see that the integrals over the circular
pieces vanish when appropriate limits are taken, the integral along $L_+$ is
the integral we are looking for, and the integral along $L_-$ differ from it 
by a phase shift. Then, it becomes
\beq
I=\int_0^\infty z^{\alpha-1} R(z)\, {\rm d}z
=\frac{2\pi i}{1-e^{2\pi i\alpha}}\sum
{\rm res}_{z_k}\left(z^{\alpha-1}R(z)\right),
\eeq
where ${\rm res}_{z_k}$ denotes the residue of a function at a pole $z_k$ and
sum goes over all poles of function $R(z)$. Applying this method, we obtain
\beq
{\cal I}=\frac{(A^2-4)}{16}\frac{\pi}{\cos(\pi/A)}
\left(B^{1/A}+B^{-1/A}\right).
\label{I}
\eeq

The second integral, ${\cal J}$, consists of three terms, due to the logarithm
of a product. The first of them is identical to $I$, i.e.,
$J_1=\int_0^\infty z^{\alpha-1} R(z) \ln A^2\, {\rm d}z=I\ln A^2$. The second
one, $J_2=\int_0^\infty z^{\alpha-1} R(z) \ln z\, {\rm d}z$, may be evaluated 
either directly via residues, as above, because $\ln z$ does not affect the
convergence of the integral, or by noticing that
$\int_0^\infty z^{\alpha-1} R(z) \ln z\, {\rm d}z=\pd{}{\alpha}{} 
\int_0^\infty z^{\alpha-1} R(z) \, {\rm d}z$ and using the previous result.
The evaluation of the last integral,
$J_3=\int_0^\infty z^{\alpha-1} (1+z)^{-3} \ln (1+z)\, {\rm d}z$, is trickier
because there is another branching point, $z=-1$, which coincide with the pole
of the rational function. We modify it as follows,
\bea
J_3&=&\int_0^\infty\frac{z^{\alpha-1}\ln (1+z)}{(1+z)^{3}}\, {\rm d}z
=-\left.\pd{}{\beta}{}\int_0^\infty\frac{z^{\alpha-1}}{(1+z)^\beta}\, {\rm d}z
\right|_{\beta=3} \nonumber \\
&=&-\left.\pd{}{\beta}{}\,B(\alpha,\beta-\alpha)\right|_{\beta=3}
=B(\alpha,3-\alpha)\left[\psi(3)-\psi(3-\alpha)\right],
\eea
where $B(m,n)=\Gamma(m)\Gamma(n)/\Gamma(m+n)$ is the beta-function,
$\psi(x)=\Gamma'(x)/\Gamma(x)$ is the digamma function, and $\Gamma(x)$ is the
gamma-function. Using the identity $\psi(x+1)=\psi(x)+1/x$ to express
$\psi(3)$ and combining all terms together, we arrive at the following result,
\bea
{\cal J}=2{\cal I}\left(\ln A-\frac{3}{2}+\gamma\right)
&+&\frac{\pi}{16\cos(\pi/A)}\left[8A-\left(A^2-4\right)\,\pi\tan(\pi/A)\right]
\left(B^{1/A}-B^{-1/A}\right) \nonumber \\
&+&\frac{\pi(A^2-4)}{8\cos(\pi/A)}
\left[B^{1/A}\,\psi\left(\frac{3}{2}+\frac{1}{A}\right)
+B^{-1/A}\,\psi\left(\frac{3}{2}-\frac{1}{A}\right)\right] ,
\label{J}
\eea
where $\gamma\equiv-\psi(1)=0.57721\ldots$ is Euler's constant. This
equation may be further simplified using the identities: 
$\psi(x+1)=\psi(x)+1/x$ and 
$\psi\left(\frac{1}{2}+x\right)-\psi\left(\frac{1}{2}-x\right)=\pi\tan(\pi x)$.
We arrive at the following result,
\beq
{\cal J}={\cal I}\left[\ln A^2-3+2\gamma+\frac{4A^2}{A^2-4}
-\pi\tan\frac{\pi}{A}+2\,\psi\!\left(\frac{1}{2}+\frac{1}{A}\right)\right] .
\eeq

The last integral, ${\cal K}$, is similar to ${\cal I}$. It is evaluated via 
the residue theorem to yield
\beq
{\cal K}=\frac{\pi(A^2-1)}{12\sin(\pi/A)}\left(B^{1/A}+B^{-1/A}\right).
\eeq

\section{Gravitational potential of a truncated, nonrotating system \label{A3} }

Here we analytically calculate the gravitational potential of the
nonrotating, $A=2$, system truncated along the cutoff equidensity (and, hence,
equipotential) surface with density $\rho=\rho_c$. Note, since we cut 
the system along the equipotential surface, the potential produced by the 
``inside'' mass in the outer space and the potential inside the cavity due 
to all ``outside'' mass are of equal magnitude and add up to a constant. 
The direct evaluation of the gravitational potential via the volume integration 
involves elliptic integrals and turns out to be cumbersome. We use a different 
approach.

Let us assume, without loss of generality, that $B>1$. Then, the ellipsoids 
are shifted upwards, as follows from Eq.\ (\ref{ellips}) and plotted in 
Fig.\ \ref{f:ellips}. We label each ellipsoid with the quantity 
\beq
\Delta=a\,\epsilon=\Delta_0/\sqrt{\rho},
\eeq
where $a$ is the major axis, $\epsilon$ is the eccentricity, and 
$\Delta_0=-\sqrt{c_s^2/8\pi G}\left(B^{1/2}-B^{-1/2}\right)$. We now consider
two such confocal ellipsoids; the quantities referred to a larger one are 
denoted by the ``prime'', as shown in Fig.\ \ref{f:ellips}. Let us now fill the 
space between the two ellipsoids with matter of a homogeneous density $\rho$.
The gravitational potential in the empty space {\em inside} the smaller
ellipsoid is equal to the potential inside the large homogeneous ellipsoid,
$\Phi'_{\rm int}({\bf x})$, less the potential inside the small one, 
$\Phi_{\rm int}({\bf x})$, having the same density. The gravitational
potential in the interior of a homogeneous prolate ellipsoid centered 
at the origin is known (\cite{BT}):
\beq
\Phi_{\rm int}({\bf x})=-\pi G\rho\left(I\,b^2-\sum_{i=1}^3 A_i\,x_i^2\right),
\eeq
where $b=a\sqrt{1-\epsilon^2}$ is the minor axis, ${\bf x}=(x,y,z)$, and
\bml
\bea
I&=&\frac{\sqrt{1-\epsilon^2}}{\epsilon}
\ln\!\left(\frac{1+\epsilon}{1-\epsilon}\right), \\
A_1=A_2&=&\frac{1-\epsilon^2}{\epsilon^2}
\left[\frac{1}{1-\epsilon^2}-\frac{1}{2\epsilon}
\ln\!\left(\frac{1+\epsilon}{1-\epsilon}\right)\right], \\
A_3&=&2\;\frac{1-\epsilon^2}{\epsilon^2}\left[
\frac{1}{2\epsilon}\ln\!\left(\frac{1+\epsilon}{1-\epsilon}\right)-1\right].
\eea
\label{IAAA}
\eml
Taking into account that our coordinate system is shifted through $\Delta<0$, 
we write, $\sum A_i x_i^2=A_1 R^2+A_3(z+\Delta)^2$. The potential inside the 
shell bounded by the surfaces $\Delta'$ and $\Delta$ is, thus,
\beq
\Phi_{\rm inside}=\Phi'_{\rm int}-\Phi_{\rm int}
=-\pi G\rho\left[\left(I\;\frac{1-\epsilon^2}{\epsilon^2}-A_3\right)
\left(\Delta'^2-\Delta^2\right)-2A_3 z \left(\Delta'-\Delta\right)\right] .
\eeq
Making $\Delta'$ infinitesimally close to $\Delta$, we obtain the contribution
to the potential due to the infinitely thin shell of ``thickness''
$\delta\Delta=\Delta'-\Delta$ and density $\rho=(\Delta_0/\Delta)^2$:
\beq
\delta\Phi_{\rm inside}=-2\pi G\left(\frac{\Delta_0}{\Delta}\right)^2
\left[\left(I\;\frac{1-\epsilon^2}{\epsilon^2}-A_3\right) \Delta\;\delta\Delta
-A_3 z\;\delta\Delta\right].
\label{phi-shell}
\eeq

The gravitational potential inside the truncated system extending from the
cutoff surface $\Delta_c=\Delta_0/\sqrt{\rho_c}$ to infinity is simply the
integral over all thin shells with $\Delta>\Delta_c$, i.e., 
$\Phi_{\rm cav}=\int\delta\Phi_{\rm inside}$. The first term in Eq.\
(\ref{phi-shell}) evaluates to a constant, which may be absorbed into $\Phi_0$, 
the zero-level of the potential. The second term yields the coordinate
dependent contribution we are looking for,
\beq
\Phi_{\rm cav}({\bf x})=-z\,\sqrt{\pi G c_s^2/2}\left(B^{1/2}-B^{-1/2}\right)
A_3(\epsilon)\sqrt{\rho_c} ,
\label{phi-cav}
\eeq
where $\epsilon=(1-B)/(1+B)$ and $A_3(\epsilon)=A_3(-\epsilon)$ is
given by Eq.\ (\ref{IAAA}). We emphasize that the potential is {\em linear} 
in the vertical coordinate, $z$, which corresponds to the homogeneous 
gravitational field. The strength of this field is determined by $B$ and the
cutoff density, $\rho_c$, only. The gravitational acceleration, 
${\bf g}=-\nabla \Phi_{\rm cav}$, is upwards for $B>1$ and downwards for $B<1$.

\end{appendix}

\figcaption[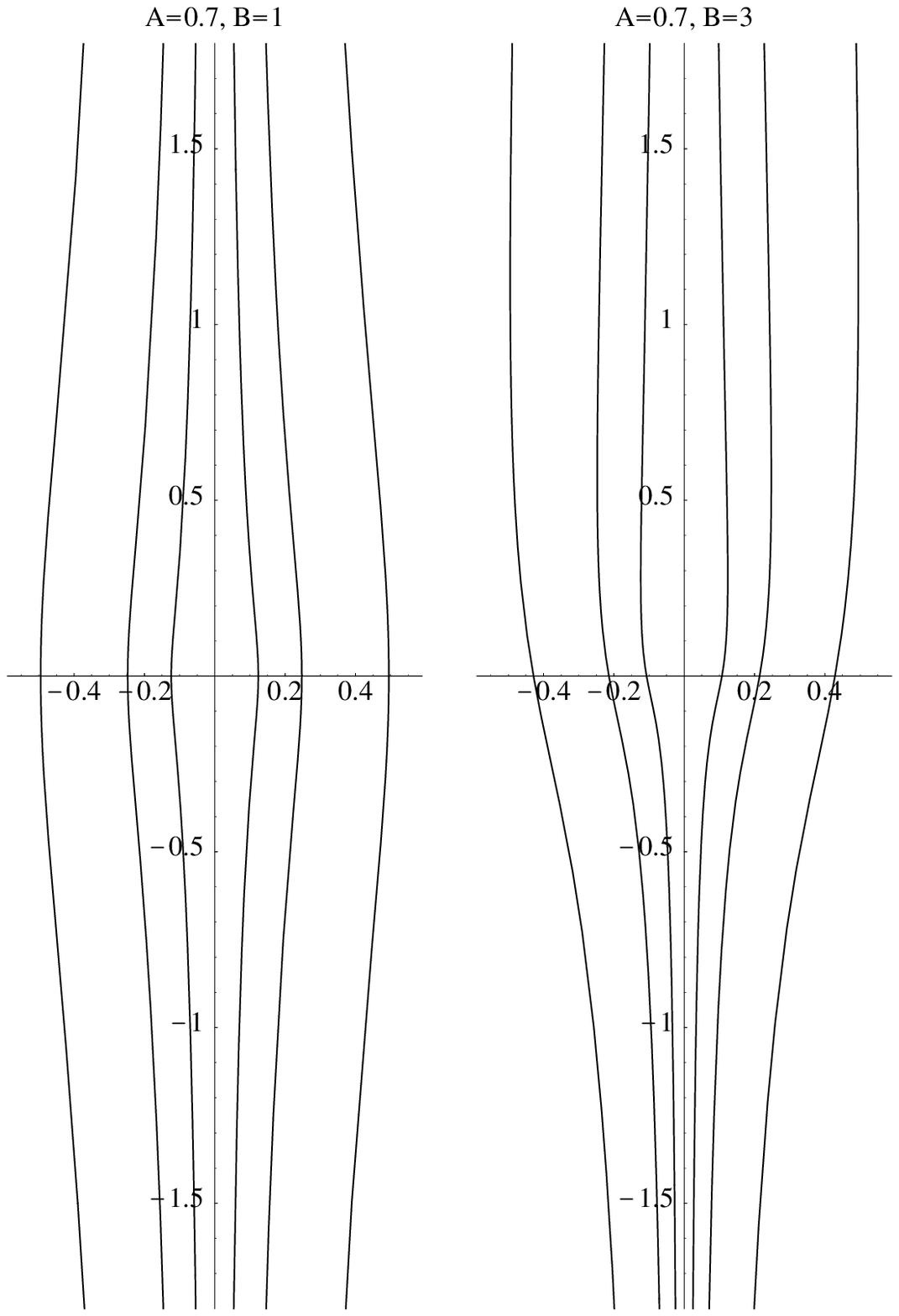]{Iso-density contours for 
$A=0.7,\, B=1 \textrm{ and } B=3$. The horizontal axis corresponds to $R$ and
the vertical axis to $z$.
\label{f:shape-a} } 

\figcaption[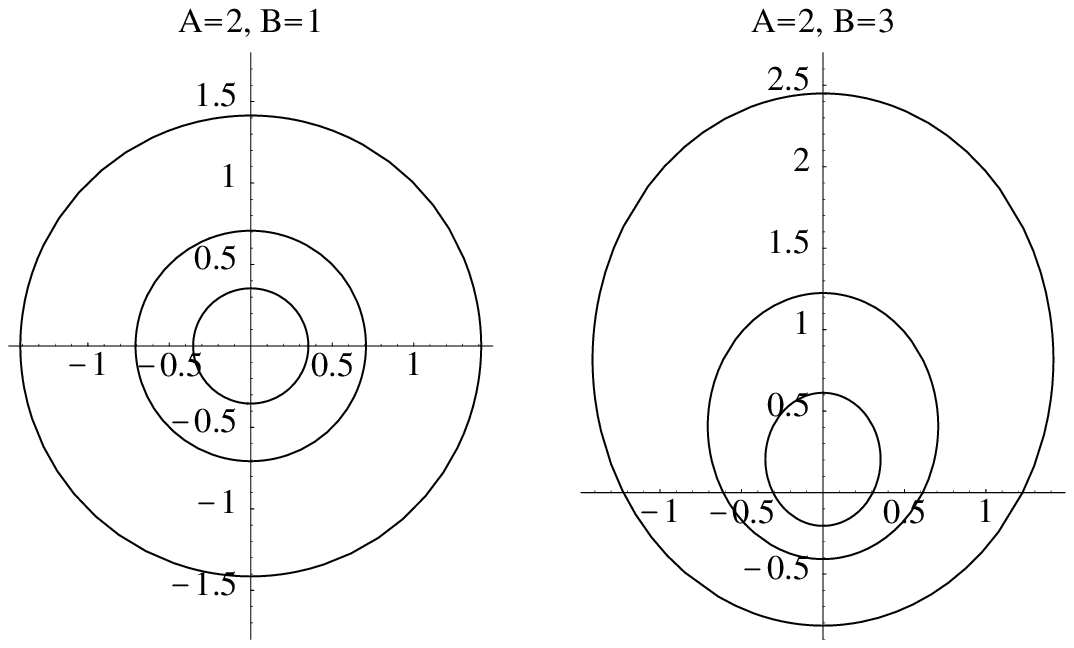]{Same as in Fig.\ \ref{f:shape-a} for
$A=2,\, B=1 \textrm{ and } B=3$. \label{f:shape-b} }

\figcaption[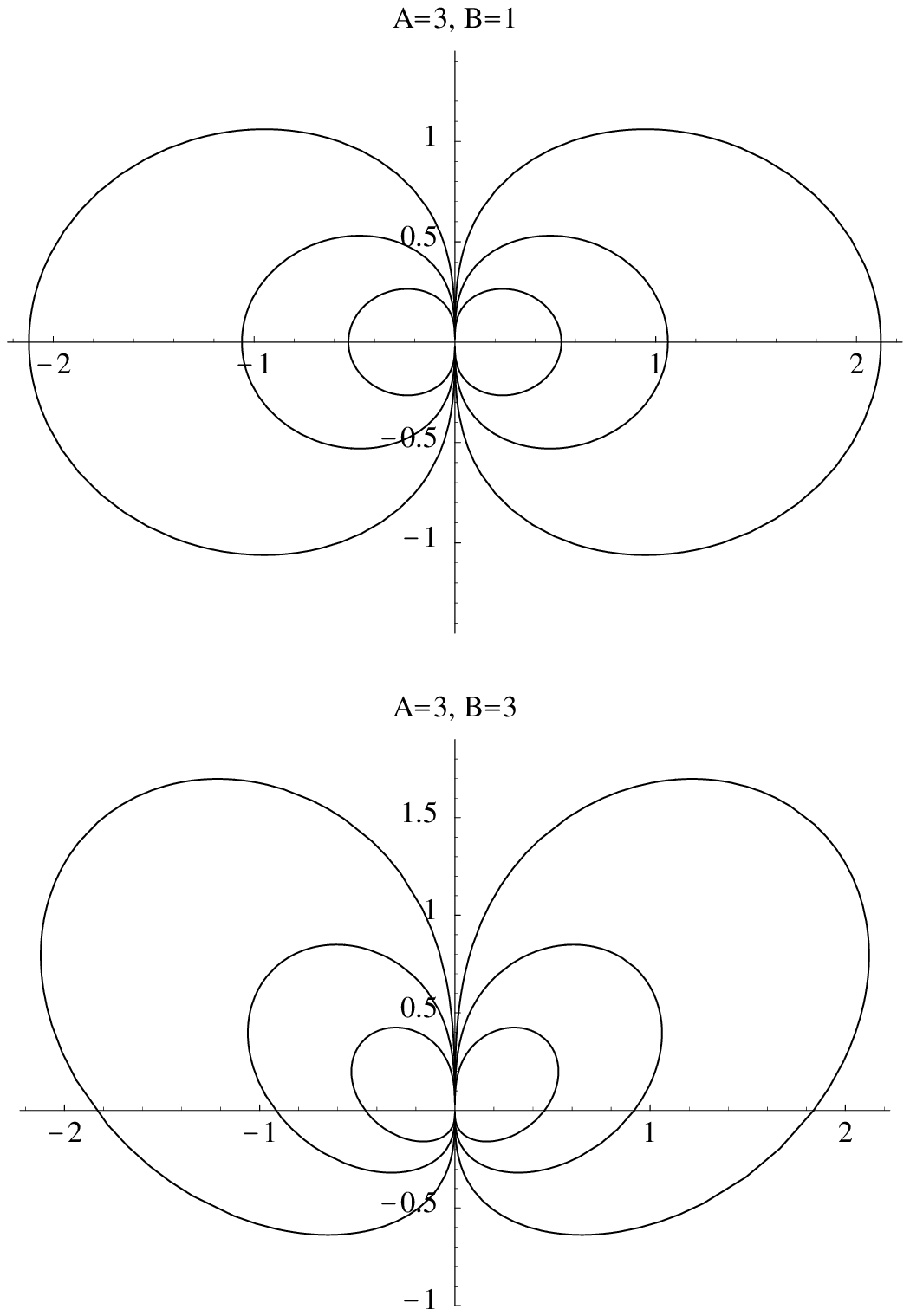]{Same as in Fig.\ \ref{f:shape-a} for
$A=3,\, B=1 \textrm{ and } B=3$. \label{f:shape-c} } 

\figcaption[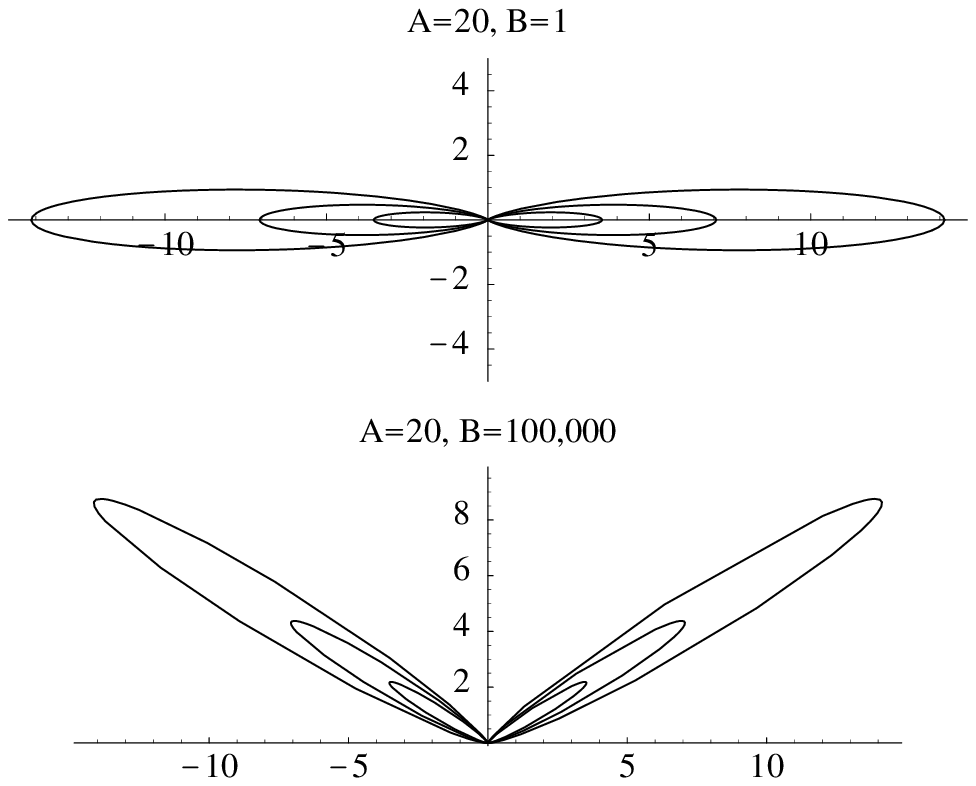]{Same as in Fig.\ \ref{f:shape-a} for
$A=20,\, B=1 \textrm{ and } B=10^5$. \label{f:shape-d} }

\figcaption[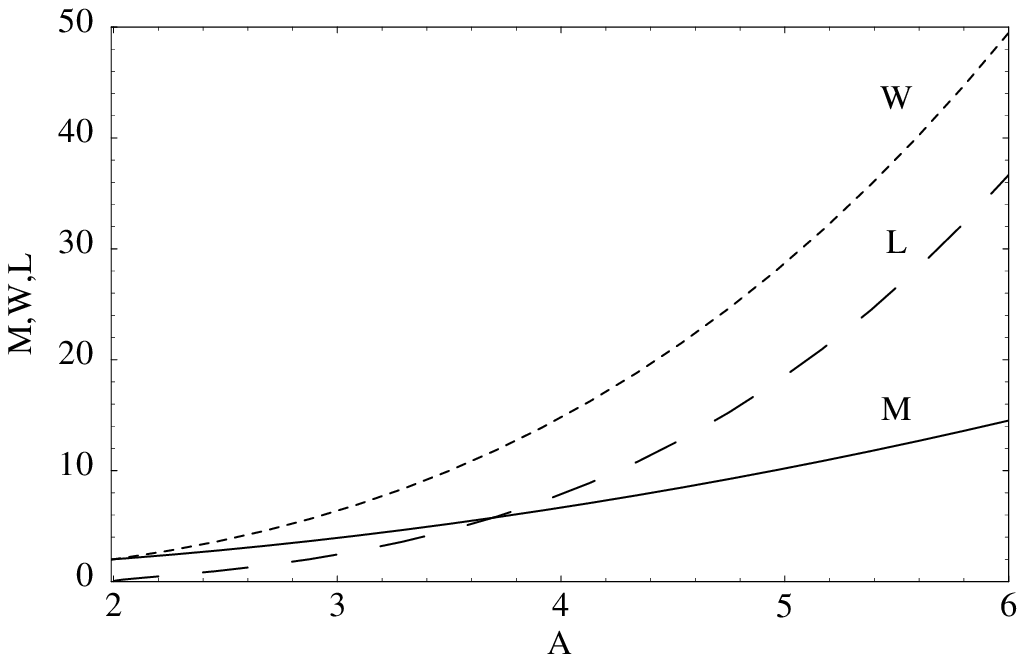]{The total mass $M$ (solid curve), gravitational energy
$W$ (short-dashed curve), and angular momentum $L_z$ (long-dashed 
curve), of a finite system vs. $A$. All quantities are in arbitrary units.
\label{f:MWL} }

\figcaption[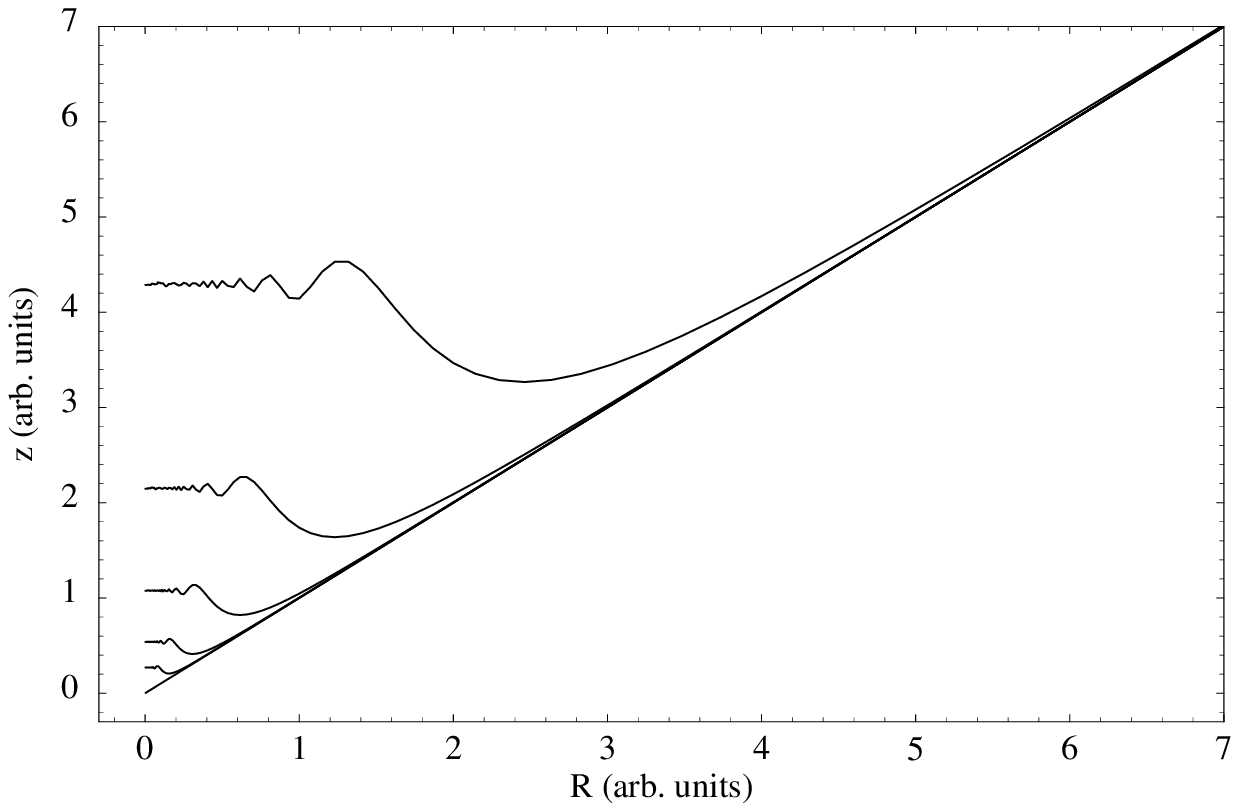]{The evolution of a free axially symmetric 
``conical disk'' as calculated by Toomre (1999, private communication).
Radial profiles of density in cylindrical coordinates 
are shown at times: $t=0$ (straight line) and (from botton 
to top) $t=0.25,\ 0.5,\ 1,\ 2,\ 4$. The coordinates and time are in arbitrary 
units.
\label{fig:toomre} }

\figcaption[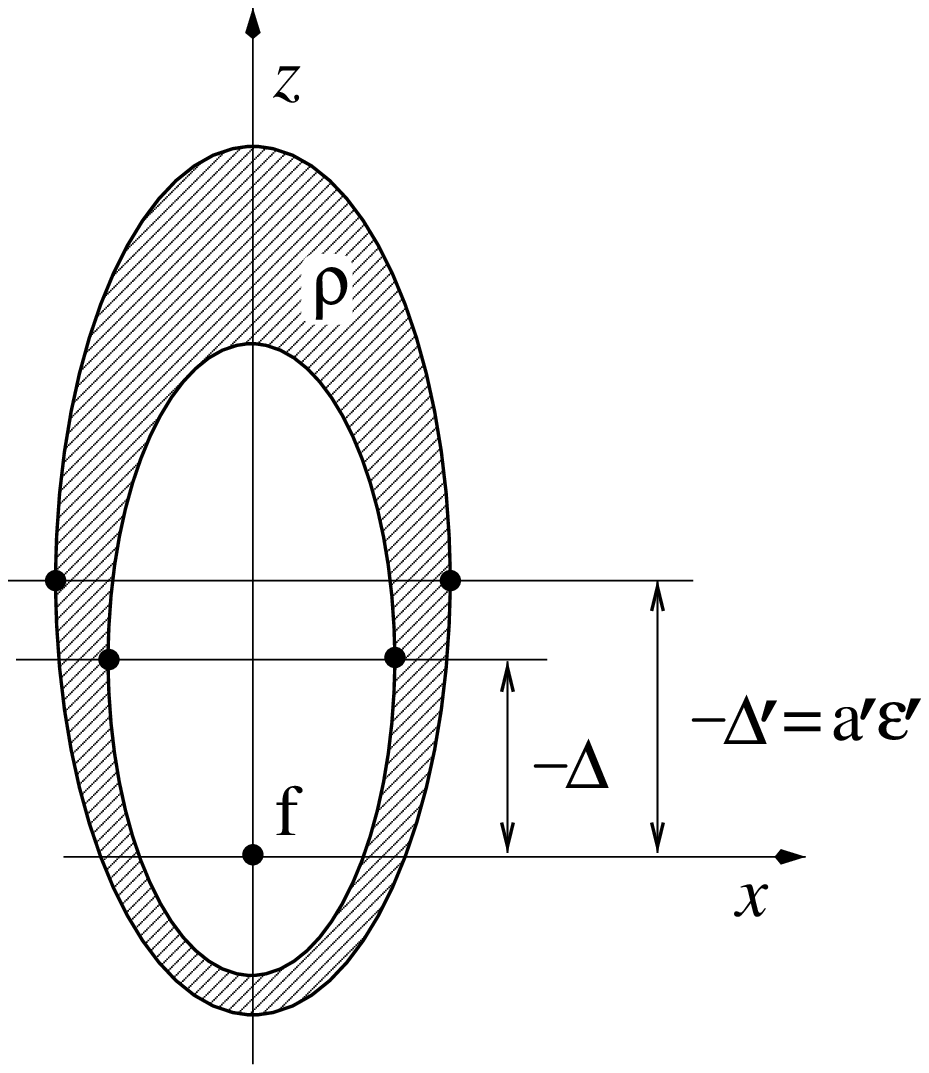]{Two confocal ellipsoids, used to 
calculate the potential of a nonrotating truncated system.
\label{f:ellips} }

\rem{
\newpage\plotone{shape-a.eps}~Fig.~\ref{f:shape-a}\vskip3cm
\newpage\plotone{shape-b.eps}\vskip1cm Fig. \ref{f:shape-b}\vskip3cm
\newpage\plotone{shape-c.eps}~Fig.~\ref{f:shape-c}\vskip3cm
\newpage\plotone{shape-d.eps}\vskip1cm Fig. \ref{f:shape-d}\vskip3cm
\newpage\plotone{mwl.eps}\vskip1cm Fig. \ref{f:MWL}\vskip3cm
\newpage\plotone{toomre.eps}\vskip1cm Fig. \ref{fig:toomre}\vskip3cm
\newpage\plotone{ellips.eps}\vskip1cm Fig. \ref{f:ellips}\vskip3cm
}

\newpage
\plottwo{shape-a.eps}{shape-c.eps}\vskip0.3cm 
	Figs. \ref{f:shape-a},\ref{f:shape-c}\vskip0.6cm
\plottwo{shape-b.eps}{shape-d.eps}\vskip0.3cm 
	Figs. \ref{f:shape-b},\ref{f:shape-d}\vskip0.6cm
\plottwo{mwl.eps}{toomre.eps}\vskip0.5cm 
	Figs. \ref{f:MWL},\ref{fig:toomre}\vskip0.6cm
\plottwo{ellips.eps}{empty.eps}\vskip0.5cm Fig. \ref{f:ellips}

\end{document}